\documentclass[prd,aps,nofootinbib,floatfix,10pt]{revtex4}

\usepackage{amsmath,graphicx,epsfig,amssymb,dsfont,mathtools,}
\usepackage[usenames]{color}
\usepackage{ulem} 
\usepackage{bigstrut}
\usepackage{multirow}
\usepackage{makecell}
\usepackage{verbatim}
\usepackage{amsmath,graphicx,color,epsfig}
  \allowdisplaybreaks[1]

\allowdisplaybreaks
\renewcommand{\thesubsection}{\Roman{subsection}}

\begin{document}
\title{Fully Heavy Tetraquark ${bb \bar c \bar c}$: Lifetimes and Weak Decays }
\author{Gang Li~$^1$, Xiao-Feng Wang~$^1$, and
Ye Xing$^2$~\footnote{Email:xingye\_guang@sjtu.edu.cn}
}
\affiliation{
$^1$ School of Physics and Engineering, Qufu Normal University, Qufu 273165, China }
\affiliation{
 $^2$ INPAC,  SKLPPC, MOE Key Laboratory for Particle Physics, School of Physics and Astronomy, \\ Shanghai Jiao Tong University, Shanghai  200240, China}

\begin{abstract}
We study  the lifetime and   weak decays of the full-heavy S-wave $0^+$ tetraquark $T^{\{bb\}}_{\{\bar c\bar c\}}$.   Using the operator product expansion rooted in heavy quark expansion, we find  a rather  short lifetime, at the order  $(0.1-0.3)\times 10^{-12}s$ depending on the inputs. With the flavor SU(3)  symmetry, we then construct   the effective Hamiltonian at the hadron level,  and derive   relations between decay widths of different channels. According to the electro-weak effective operators,   we classify different decay modes, and make  a collection of  the golden channels, such as $T_{\{\bar c \bar c\}}^{\{bb\}}\to B^- K^0 B_c^-$ for the charm quark decay and $T_{\{\bar c \bar c\}}^{\{bb\}}\to B^-D^-$ for the bottom quark decay.
Our results for the lifetime and golden channels are helpful to search for the fully-heavy tetraquark in future experiments.
\end{abstract}

\maketitle

\section{Introduction}

In the past decades, quark model has achieved great successes in the hadron spectroscopy study. In addition to the quark-anti-quark assignment for a meson and three-quark interpretation of  a baryon, it   allows the existence of non-standard  exotic states~\cite{Choi:2003ue,Belle:2011aa,Ablikim:2013mio,Ablikim:2013emm,Choi:2007wga,Aaij:2015tga}.
Since the observation of $X(3872)$ in 2003~\cite{Choi:2003ue}, many exotic candidates have been   announced on the experimental side in the heavy quarkonium sector
in various processes~\cite{Tanabashi:2018oca}. Charged heavy quarkoniumlike states $Z_c(3900)^{\pm}$, $Z_c(4020)^{\pm}$, $Z_b(10610)^{\pm}$, and $Z_b(10650)^{\pm}$ observed by BES-III and Belle collaborations~\cite{Belle:2011aa,Ablikim:2013mio,Ablikim:2013emm} have already experimentally established as being exotic, since they contain at least two quarks and two antiquarks with the hidden $Q\bar Q$. Until now, extensive
theoretical studies have been carried out to explore their internal  structures, production  and decay behaviors~\cite{Guo:2013sya,Cleven:2013sq,Guo:2013ufa,Liu:2016xly,Li:2014pfa,Li:2013xia,Guo:2014sca,Guo:2014ppa,Guo:2014iya,Chen:2016mjn,Wang:2013kra,Li:2012as,Li:2014uia,Albaladejo:2017blx,Liu:2013vfa,Guo:2013zbw,Voloshin:2013ez,Voloshin:2011qa,Chen:2011pv,Li:2013yla,Chen:2011pu,Chen:2012yr,Bondar:2011ev,Li:2015uwa,Chen:2013bha,Wu:2016ypc}.  Most of the established  states tend to contain a pair of heavy quark, and thus the discovery of   exotic states of new categories  will be valuable. Fully-heavy four-quark state with no light quark degrees of freedom  is of this type and might be an   ideal  probe   to study the interplay between perturbative QCD and non-perturbative QCD.

Generally speaking, more heavy quarks correspond to a larger mass. For instance, there have been some phenomenological studies to determine the mass and the spectrum properties of the fully-heavy tetraquark  ${b\bar c}{b\bar c}$, including the constituent quark and diquark model~\cite{Bai:2016int,Karliner:2016zzc}, chiral quark model~\cite{Wu:2016vtq}, nonrelativistic effective field theory(NREFT)~\cite{Anwar:2017toa}, and QCD sum rules~\cite{Chen:2016jxd,Wang:2017jtz}. In Ref.~\cite{Anwar:2017toa}, the authors  utilize the NREFT to determine the mass with the upper bound as $12.58$ GeV,  consistent with the mass  calculated in the chiral quark model~\cite{Wu:2016vtq}.   Despite of  these studies, it is still not conclusive  that  whether the ${b\bar c}{b\bar c}$  (or its charge conjugate ${c\bar b}{c\bar b}$) is above or below the $B_c B_c$ threshold. It is likely that  the  ${b\bar c}{b\bar c}$ lies below the threshold of  the $B_c B_c$ pair, which means that such a   state is stable against the strong interaction. In this case,  the dominant decay modes would  be induced by weak interaction. In a diquark-diquark model~\cite{Jaffe:2003sg}, the S-wave fully-heavy tetraquark state  ${b\bar c}{b\bar c}$ can form $0^+$ and $2^+$. In this paper we will mainly focus  on the lowest lying state $0^+$, which might be assigned as  a weakly-coupled state.

In this paper, we will first use the operator product expansion (OPE) technique and calculate  the  lifetime of the S-wave $0^+$ ${b\bar c}{b\bar c}$.
The light flavor SU(3) symmetry is a useful tools to analyze weak decays of a heavy quark, and  has  been successfully applied to the meson or baryon system~\cite{Savage:1989ub,Gronau:1995hm,He:1998rq,Chiang:2004nm,Li:2007bh,Wang:2009azc,Cheng:2011qh,Hsiao:2015iiu,Lu:2016ogy,He:2016xvd,Wang:2017vnc,Wang:2017mqp,Wang:2017azm,Shi:2017dto,Wang:2018utj,He:2018php}. Though the SU(3) breaking effects in charm quark transition might  be sizable, the results from the flavor symmetry can describe the experimental  data in a global viewpoint. To be more explicit,  one can write down the Hamiltonian at the hadron level  with   hadron fields and transition operators. Some limited amount of input  parameters will be introduced  to describe the non-perturbative transitions. With the SU(3) amplitudes,  one can obtain   relations between  decay widths of different processes,  which can be examined  in experiment. Such an analysis is also helpful to identify the decay modes that will be mostly useful to discover the fully-heavy tetraquark state.

The rest of this paper is organized as follows. In Sec.~\ref{sec:particle_multiplet}, we give the particle multiplets under the SU(3) symmetry. Section III is devoted to calculate  the lifetime of the tetraquark state using the OPE.  In Section IV, we discuss the weak decays of many-body final states, including mesonic two-body or three-body decays and baryonic two-body decays. In section V, we present a collection of the golden channels. Finally, we provide a short summary.

\section{Particle Multiplets in SU(3)}
\label{sec:particle_multiplet}
The tetraquark with the quark constituents ${b\bar c}{b\bar c}$ does not contain any light quark and thus is an SU(3)   singlet.  Recalling that diquark $[QQ]$ or $[qq]$ live in $\bf{A}_{color} \otimes \bf{S}_{flavor}\otimes \bf{S}_{spin}$ spaces, with $A$ and $S$ representing the symmetry and anti-symmetry representation respectively, we find the allowed spin quantum numbers are $\bf1\otimes\bf1=\bf0\oplus\bf2$. In this paper,  we will mainly  focus on the lowest lying  state with $J^P=0^+$, which  is abbreviated  as $T_{\{\bar c \bar c\}}^{\{bb\}}$.

\begin{figure}
  \centering
  \includegraphics[width=0.75\columnwidth]{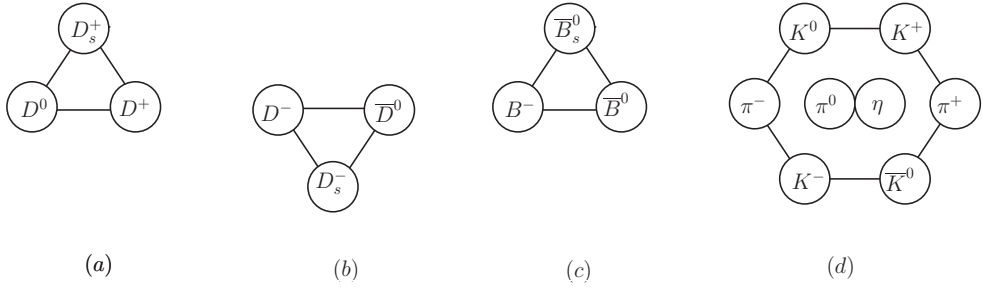}\\
  \caption{The weight diagrams for the anti-charmed meson triplet, charmed meson anti-triplet, bottom meson triplet and light meson octet}\label{fig:multiplet-mesons}
\end{figure}

\begin{figure}
  \centering
  \includegraphics[width=0.9\columnwidth]{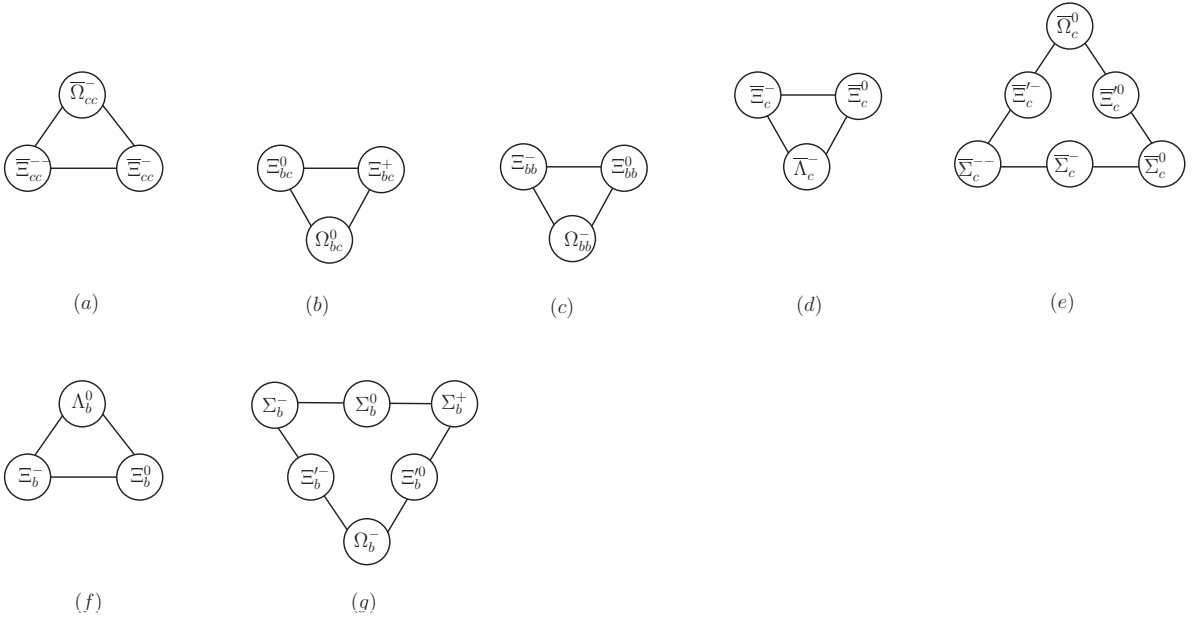}\\
  \caption{The weight diagrams for the doubly heavy baryon are given in (a,b,c), which anti-triplet $\overline F_{c c}$ to be (a), triplet $F_{bc}$ to be (b), or triplet $F_{bb}$ to be (c). The singly anti-charm baryon multiplets are $\overline F_{ c3}, \overline F_{c\bar6} $ shown in (d,e), and the singly bottom baryon multiplets are given in (f,g) signed as $F_{b\bar3},F_{b6} $.}\label{fig:multiplet-baryons}
\end{figure}

In the baryon sector,  we give the SU(3) representations for baryons with different charm quantum numbers ($C$) or bottom quantum numbers ($B$) as follows.
The triply heavy baryon with $C=-3$ denoted  as $\overline F_{ccc}$ can form an SU(3) singlet ${\overline \Omega}_{ccc}^{\, --}$. Baryons with doubly heavy quarks(i.e. $C=-2$, $B,C=1$, $B=2$) are supposed to be an anti-triplet(triplet) given as
\begin{eqnarray}
 (\overline F_{ c c}^T)_i  = \left(\begin{array}{c}  \overline \Xi^{--}_{cc}(\bar c\bar c\bar u) \\ \overline \Xi^-_{cc}(\bar c\bar c\bar d) \\ \overline \Omega^-_{cc}(\bar c\bar c\bar s)
\end{array}\right)\,,\;\;
  F_{bc}^i = \left(\begin{array}{c}  \Xi^+_{bc}(bcu)  \\  \Xi^0_{bc}(bcd)  \\  \Omega^0_{bc}(bcs)
\end{array}\right)\,,\;\;
  F_{bb}^i = \left(\begin{array}{c}  \Xi^0_{bb}(bbu)  \\  \Xi^-_{bb}(bbd)  \\  \Omega^-_{bb}(bbs)
\end{array}\right).
\end{eqnarray}
Consistently, the singly heavy baryons with  $C=-1$($B=1$) are expected  to form a triplet(anti-triplet) and a anti-sextet(sextet) as~\cite{Xing:2018bqt}
\begin{eqnarray}
 (\overline F_{\bf{c 3}})_{[ij]}= \left(\begin{array}{ccc} 0 & \overline \Lambda_{c}^-  &  \overline \Xi_{c}^-  \\ -\overline \Lambda_{c}^- & 0 & \overline \Xi_{c}^0 \\ -\overline \Xi_{c}^-   &  -\overline \Xi_{c}^0  & 0
  \end{array} \right), \;\;
 (\overline F_{\bf{c\bar{6}}})_{\{ij\}} = \left(\begin{array}{ccc} \overline \Sigma_{c}^{--} &  \frac{1}{\sqrt{2}}\overline \Sigma_{c}^-   & \frac{1}{\sqrt{2}} \overline \Xi_{c}^{\prime-}\\
  \frac{1}{\sqrt{2}}\overline \Sigma_{c}^-& \overline \Sigma_{c}^{0} & \frac{1}{\sqrt{2}} \overline \Xi_{c}^{\prime0} \\
  \frac{1}{\sqrt{2}} \overline \Xi_{c}^{\prime-}   &  \frac{1}{\sqrt{2}} \overline \Xi_{c}^{\prime0}  &
   \overline \Omega_{c}^0
  \end{array} \right)\,\nonumber\\
 (F_{\bf{b \bar3}})^{[ij]}= \left(\begin{array}{ccc} 0 & \Lambda_{b}^0  &  \Xi_{b}^0  \\ -\Lambda_{b}^0 & 0 & \Xi_{b}^- \\ -\Xi_{b}^0   &  -\Xi_{b}^-  & 0
  \end{array} \right), \;\;
 (F_{\bf{b6}})^{\{ij\}} = \left(\begin{array}{ccc} \Sigma_{b}^{+} &  \frac{1}{\sqrt{2}}\Sigma_{b}^0   & \frac{1}{\sqrt{2}} \Xi_{b}^{\prime0}\\
  \frac{1}{\sqrt{2}}\Sigma_{b}^0&  \Sigma_{b}^{-} & \frac{1}{\sqrt{2}} \Xi_{b}^{\prime-} \\
  \frac{1}{\sqrt{2}} \Xi_{b}^{\prime0}   &  \frac{1}{\sqrt{2}}  \Xi_{b}^{\prime-}  &  \Omega_{b}^-
  \end{array} \right)\,.
\end{eqnarray}
In the meson sector, singly heavy mesons form an SU(3) triplet or anti-triplet, while the light mesons form an octet plus a flavor singlet. These multiplets can be written as
\begin{eqnarray}
 M_{8}=\begin{pmatrix}
 \frac{\pi^0}{\sqrt{2}}+\frac{\eta}{\sqrt{6}}
 &\pi^+ & K^+\\
 \pi^-&-\frac{\pi^0}{\sqrt{2}}+\frac{\eta}{\sqrt{6}}&{K^0}\\
 K^-&\bar K^0 &-2\frac{\eta}{\sqrt{6}}
 \end{pmatrix},
B_i^{T}  = \left(\begin{array}{c}   B^-  \\  \overline B^0  \\  \overline B^0_s
\end{array}\right)\,,\;\;
D_i^T = \left(\begin{array}{c}  D^0  \\  D^+  \\  D^+_s
\end{array}\right)\,,\;\;
\overline D^i= \left(\begin{array}{c} \overline D^0  \\  D^-  \\  D^-_s
\end{array}\right).
\end{eqnarray}
The weight diagrams of the multiplets  are given in Fig.~\ref{fig:multiplet-mesons} and  Fig.~\ref{fig:multiplet-baryons}.

\section{LIFETIME}
\label{sec:particle_lifetime}
In this section we will discuss the lifetime of $T^{\{bb\}}_{\{\bar c\bar c\}}$ using the OPE~\cite{Lenz:2014jha,Ali:2018xfq}. The decay width of $T^{\{bb\}}_{\{\bar c\bar c\}}\to X$ are as follows:
\begin{eqnarray}
\Gamma(T^{\{bb\}}_{\{\bar c\bar c\}}\to X)=\frac{1}{2m_T} \sum_X \int \prod_i \Big[ \frac{d^3 \overrightarrow{p}_i}{(2\pi)^3 2E_i}\Big] (2\pi)^4 \delta^4(p_T-\sum_i p_i) \sum_{\lambda}|\langle X|\mathcal{H}|T^{\{bb\}}_{\{\bar c\bar c\}}\rangle |^2,
\end{eqnarray}
where $m_T$, $p_T^{\mu}$, and $\lambda$ are the mass, four-momentum and spin of $T^{\{bb\}}_{\{\bar c\bar c\}}$, respectively.
The   electro-weak effective Hamiltonian $\mathcal{H}_{eff}^{ew}$ is  given as
\begin{eqnarray}
\mathcal{H}_{eff}^{ew}=\frac{G_F}{\sqrt{2}} \Big[ \sum_{q=u,c} V_c^q(C_1 O_1^q+C_2 O_2^q)-V_p \sum_{j=3} C_j O_j\Big]
\end{eqnarray}
here, $C_i$ and $O_i$ are Wilson coefficients and operators. $V$s are the combinations of Cabibbo-Kobayashi-Maskawa(CKM) elements.
Using the optical theorem, the total decay width of $\Gamma(T^{\{bb\}}_{\{\bar c\bar c\}}\to X)$ can be rewritten as
\begin{eqnarray}
\Gamma(T^{\{bb\}}_{\{\bar c\bar c\}}\to X)=\frac{1}{2m_T}\sum_{\lambda}\langle T^{\{bb\}}_{\{\bar c\bar c\}}|\mathcal{T}|T^{\{bb\}}_{\{\bar c\bar c\}}\rangle\\
\mathcal{T}=Im\ i \int d^4x T\{ \mathcal{H}_{eff}(x) \mathcal{H}_{eff}(0)\}
\end{eqnarray}
In the heavy quark expansion (HQE),  the transition operators up to dimension 6 contribute:
\begin{eqnarray}
\mathcal{T}=\sum_{Q=b,c}\frac{G_F^2m_Q^5}{192\pi^3} |V_{CKM}|^2 \Big[c_{3,Q}(\bar QQ)+\frac{c_{5,Q}}{m_Q^2}(\bar Q g_s \sigma_{\mu\nu}G^{\mu\nu}Q)+2\frac{c_{6,Q}}{m_Q^3}(\bar Qq)_{\Gamma}(\bar q Q)_{\Gamma} \Big],
\end{eqnarray}
with $G_F$ being the Fermi constant and $V_{CKM}$ being the CKM mixing matrix. The coefficients $c_{i,Q}$ are the perturbative short-distance coefficients.
The contribution to  decay width from the lowest dimension operator is  given as
\begin{eqnarray}
\Gamma(T^{\{bb\}}_{\{\bar c\bar c\}} \to X)= \sum_{Q=b,c} \frac{G_F^2m_Q^5}{192\pi^3} |V_{CKM}|^2 c_{3,Q} \frac{\langle T^{\{bb\}}_{\{\bar c\bar c\}}|\bar Q Q|T^{\{bb\}}_{\{\bar c\bar c\}}\rangle}{2m_T},
\end{eqnarray}
where  the matrix element
\begin{eqnarray}
\frac{\langle T^{\{bb\}}_{\{\bar c\bar c\}}|\bar Q Q|T^{\{bb\}}_{\{\bar c\bar c\}}\rangle}{2m_T}
\end{eqnarray}
corresponds to the bottom and charmed number in the tetraquark state. The matrix elements of the $\bar bb$ operator and the $\bar cc$ operator give the bottom-quark and charm-quark number in the $T^{\{bb\}}_{\{\bar c\bar c\}}$ tetraquark respectively given as
\begin{eqnarray}
\frac{\langle T^{\{bb\}}_{\{\bar c\bar c\}}|\bar bb|T^{\{bb\}}_{\{\bar c\bar c\}}\rangle}{2m_T}=2+\mathcal{O}(1/m_b),\; \frac{\langle T^{\{bb\}}_{\{\bar c\bar c\}}|\bar cc|T^{\{bb\}}_{\{\bar c\bar c\}}\rangle}{2m_T}=2+\mathcal{O}(1/m_c).
\end{eqnarray}

The short distance coefficients $c_{3,Q}$s have been calculated as $c_{3,b}=5.29\pm0.35,\ c_{3,c}=6.29\pm0.72$ at the leading order(LO) and $c_{3,b}=6.88\pm0.74,\ c_{3,c}=11.61\pm1.55$ at the next-to-leading order(NLO)~\cite{Lenz:2014jha}. Therefore we expect that the total decay width and lifetime of the $T^{\{bb\}}_{\{\bar c\bar c\}}$ tetraquark as
\begin{eqnarray}
&&\Gamma(T^{\{bb\}}_{\{\bar c\bar c\}})=\left\{\begin{array}{l} (2.44\pm0.23)\times 10^{-12} \ {\rm GeV} ,
\;\text{LO} \\ (3.97\pm1.50)\times 10^{-12} \ {\rm GeV}  ,\; \text{NLO} \end{array}\right. ,\\
&&\tau(T^{\{bb\}}_{\{\bar c\bar c\}})=\left\{\begin{array}{l} (0.27\pm0.02)\times 10^{-12} \ s ,
\;\text{LO} \\ (0.17\pm0.02)\times 10^{-12} \ s  ,\; \text{NLO} \end{array}\right. ,
\end{eqnarray}
where we use the heavy quark masses $m_c=1.4\ {\rm GeV}$ and $m_b=4.8\ {\rm GeV}$. The lifetime of $T^{\{bb\}}_{\{\bar c\bar c\}}$ is much smaller than that of $B_c$ meson
\begin{eqnarray}
\tau(B_c^+)=0.507\times 10^{-12}\ s,
\end{eqnarray}
and in particular, their ratio  is about  one third.

\section{Weak decays}
\renewcommand\thesubsection{(\Roman{subsection})}
\label{sec:weak_decay}
In this section, we will discuss the possible weak decay modes of the tetraquark. Usually, the $b$ and $c$ quark in tetraquark state can decay weakly. For simplicity, we will classify the decays modes by the quantities of CKM matrix elements.
\begin{itemize}
\item For the $b/c$ quark decays into lepton pair, semi-leptonic decay process, we consider the following groups.
\begin{eqnarray}
b\to c/u \ell^- \bar \nu_{\ell},\ \ \
  \bar c\to  \bar d/\bar s  \ell^-   \bar \nu_{\ell}.
\end{eqnarray}
The general electro-weak  Hamiltonian  for the  above semi-leptonic transition  can be expressed as
\begin{eqnarray}
 {\cal H}_{eff} &=& \frac{G_F}{\sqrt2} \left[V_{q'b} \bar q' \gamma^\mu(1-\gamma_5)b \bar  \ell\gamma_\mu(1-\gamma_5) \nu_{\ell}+V_{cq} \bar c  \gamma^\mu(1-\gamma_5)q \bar \ell \gamma_\mu(1-\gamma_5) \nu_{\ell}\right] +h.c.,
\end{eqnarray}
with $q'=(u,c)$, $q=(d,s)$, in which the operator of $b\to u/c \ell^- \bar \nu_{\ell}$ transition forms an SU(3) flavor triplet $H_{3}'$ or singlet, with $(H_3')^1=1$ and $(H_3')^{2,3}=0$. Furthermore, it is easy  to see  that  the $\bar c\to \bar q \ell^-\bar \nu$ transition forms a triplet $H_{  3}$, particularly $(H_{  3})_1=0,~(H_{  3})_2=V_{cd},~(H_{  3})_3=V_{cs}$.
\item  The $c$ quark non-leptonic decays are classified as
\begin{eqnarray}
\bar c\to \bar s d \bar u, \;
\bar c\to \bar u d \bar d/s \bar s, \;
\bar c\to \bar d s \bar u, \;
\end{eqnarray}
The three kinds of decays are Cabibbo allowed, singly Cabibbo suppressed, and doubly Cabibbo suppressed respectively.
Under the flavor SU(3) symmetry, the transition $\bar c \to \bar q_1  q_2 \bar q_3$  can be decomposed
 as ${\bf  \bar3}\otimes {\bf 3}\otimes {\bf
\bar3}={\bf  \bar3}\oplus {\bf  \bar3}\oplus {\bf6}\oplus {\bf \overline {15}}$.
For the Cabibbo allowed transition   $\bar c\to \bar s  d \bar u$,  the nonzero tensor components are given as
\begin{eqnarray}
(H_{ 6})_{31}^2=-(H_{6})_{13}^2=1,\;\;\;
 (H_{\overline {15}})_{31}^2= (H_{\overline {15}})_{13}^2=1.\label{eq:H3615_c_allowed}
\end{eqnarray}
For  the  singly Cabibbo suppressed transition $\bar c\to \bar u d\bar d$ and $\bar c\to \bar u  s\bar s$, the combination of tensor components are given as
\begin{eqnarray}
(H_{6})_{31}^3 =-(H_{6})_{13}^3 =(H_{ 6})_{12}^2 =-(H_{ 6})_{21}^2 =\sin(\theta_C),\nonumber\\
 (H_{\overline {15}})_{31}^3= (H_{\overline {15}})_{13}^3=-(H_{\overline {15}})_{12}^2=-(H_{\overline {15}})_{21}^2= \sin(\theta_C).\label{eq:H3615_cc_singly_suppressed}
\end{eqnarray}
while for the doubly Cabibbo suppressed transition  $\bar c\to \bar d  s \bar u$, we have
\begin{eqnarray}
(H_{ 6})_{21}^3=-(H_{ 6})_{12}^3=-\sin^2\theta_C,\;\;
 (H_{\overline {15}})_{21}^3= (H_{\overline {15}})_{12}^3=-\sin^2\theta_C. \label{eq:H3615_c_doubly_suprressed}
\end{eqnarray}
\item The b quark non-leptonic decays are classified as:
\begin{eqnarray}
b\to c\bar c d/s, \;
b\to c \bar u d/s, \;
b\to u \bar c d/s, \;
b\to q_1 \bar q_2 q_3,
\end{eqnarray}
here $q_{1,2,3}$ represent the light quark($d/s$).

The transition operator for the  $b\to c\bar c d/s$ forms  an triplet, with $(H_{  3})^{2}=V_{cd}^*,\;(H_{  3})^{3}=V_{cs}^*$. The operator of the transition $b\to c \bar u d/s$  can form an octet $\bf8$, whose nonzero composition followed as $\;(H_{{8}})^2_1 =V_{ud}^*\; ,\;(H_{{8}})^3_1 =V_{us}^*\; $. For  the transition $b\to u\bar cs$,  the operator can form an anti-symmetric ${\bf  \bar 3}$ with $(H_{\bar 3}'')^{13} =- (H_{\bar 3}'')^{31} =V_{cs}^*$ plus a symmetric ${\bf  6}$ tensors with $(H_{ 6})^{13}=(H_{6})^{31} =V_{cs}^*$. It is straightforward  to obtain the similar transition $b\to u\bar cd$ by exchanging the index $2\to 3$ and the  $V_{cs}\to V_{cd}$ in previous transition.

The charmless   transition $b \to q_1 \bar q_2 q_3$ ($q_i=d,s$) can be decomposed as $\bf3\otimes\bar {\bf3}\otimes\bf3=\bf3\oplus\bf3\oplus \bf{\bar 6}\oplus \bf{15}$, where the triplet $H_{\bf 3}$ behave as the penguin level operator. In the $\Delta S=0 (b\to d)$ decays, the nonzero components of these irreducible tensors are given as
\begin{eqnarray}
 (H_3)^2=1,\;\;\;(H_{\overline6})^{12}_1=-(H_{\overline6})^{21}_1=(H_{\overline6})^{23}_3=-(H_{\overline6})^{32}_3=1,\nonumber\\
 2(H_{15})^{12}_1= 2(H_{15})^{21}_1=-3(H_{15})^{22}_2=
 -6(H_{15})^{23}_3=-6(H_{15})^{32}_3=6.\label{eq:H3615_bd}
\end{eqnarray}
For the $\Delta S=1 (b\to s)$ decays, the nonzero entries in the irreducible tensor $H_{\bf{3}}$, $H_{\bf\overline6}$,
$H_{\bf{15}}$ can be obtained from Eq.~\eqref{eq:H3615_bd}
with the exchange $2\leftrightarrow 3$.
\end{itemize}

In the following, we will study the possible decay modes of $T^{\{bb\}}_{\{\bar c\bar c\}}$ in order.

\subsection{Semi-Leptonic $T^{\{bb\}}_{\{\bar c\bar c\}}$ decays}
\renewcommand\thesubsection{(\roman{subsection})}
\label{sec:semileptonic_decay}
\begin{figure}
  \centering
  \includegraphics[width=0.75\columnwidth]{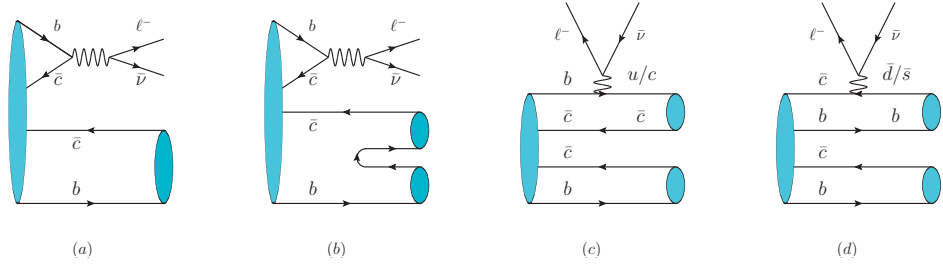}\\
  \caption{Feynman diagrams for $T^{\{bb\}}_{\{\bar c\bar c\}}$ semi-leptonic decays within $b/\bar c$ quark decay. The panel(a) represents a meson final state, while the panel(b,c,d) correspond with two mesons final states. In panel(a,b), the quarks in initial state interact by the exchange of W boson.}\label{fig:topology1}
\end{figure}


\subsubsection{$b\to c/u \ell^- \overline \nu_{\ell}$ transition }
At the hadron level, the $b\to u$ transition  can be realized by the process that $T^{\{bb\}}_{\{\bar c\bar c\}}$ decays to a anti-charmed meson plus $B_c$ meson and $\ell \overline \nu_{\ell}$.
Following the SU(3) analysis, the Hamiltonian at the hadronic level is constructed as
$a_1 T^{\{bb\}}_{\{\bar c\bar c\}} (H_3')^i D_i \overline{B}_c \bar \ell \nu,$
with the coefficient $a_{1}$ representing the non-perturbative parameter. For completeness, we give the corresponding Feynman diagram at quark level shown in Fig.~\ref{fig:topology1}.(c). It is convenient to obtain the decay amplitudes by expanding the Hamiltonian constructed above and the amplitude $\mathcal{M}(T^{\{bb\}}_{\{\bar c\bar c\}}\to \overline D^0  B_c^- l^-\bar\nu)\; =\ a_1 V_{\text{ub}}$.

For the SU(3) singlet $b\to c$ transition, the final hadrons of the many-body semileptonic decays of $T^{\{bb\}}_{\{\bar c\bar c\}}$ can be a $B_c$ meson, $B_c$ plus $J/\psi$, charmed meson plus bottom meson respectively. Consequently, the Hamiltonian at the hadron level is constructed as
\begin{eqnarray}
&\mathcal{H}_{eff}&=a_2 T^{\{bb\}}_{\{\bar c\bar c\}} \overline B_c \bar \ell \nu +a_3 T^{\{bb\}}_{\{\bar c\bar c\}} \overline B_c J/\psi \bar \ell \nu +a_4 T^{\{bb\}}_{\{\bar c\bar c\}} D_i \overline{B}^i\bar \ell \nu .
\end{eqnarray}
Feynman diagrams are shown in Fig.~\ref{fig:topology1}.(a,b). One then obtain the amplitudes of different decay channels listed in Tab.~\ref{tab:bc__btoc_lv}, from which we derive that the simple relations between different decay widths as:
$\Gamma(T^{\{bb\}}_{\{\bar c\bar c\}}\to \overline D^0  B^- l^-\bar\nu)=
\Gamma(T^{\{bb\}}_{\{\bar c\bar c\}}\to D^-  \overline B^0 l^-\bar\nu)=
\Gamma(T^{\{bb\}}_{\{\bar c\bar c\}}\to  D^-_s  \overline B^0_s l^-\bar\nu)$.
\begin{table}
\caption{Amplitudes for tetraquark $T^{\{bb\}}_{\{\bar c\bar c\}}$  decays into two mesons and three mesons for the transition $b\to c\ell^- \bar \nu$. }\label{tab:bc__btoc_lv}\begin{tabular}{|c|c|c|c|c|c|c|c}\hline\hline
channel & amplitude &channel & amplitude\\\hline
$T^{\{bb\}}_{\{\bar c\bar c\}}\to B_c l^-\bar\nu $ & $ a_2 V_{cb}$&
$T^{\{bb\}}_{\{\bar c\bar c\}}\to B_c J/\psi l^-\bar\nu $ & $ a_3V_{cb}$\\\hline
$T^{\{bb\}}_{\{\bar c\bar c\}}\to \overline D^0  B^- l^-\bar\nu $ & $ a_4V_{cb}$&
$T^{\{bb\}}_{\{\bar c\bar c\}}\to D^-  \overline B^0 l^-\bar\nu $ & $ a_4V_{cb}$\\\hline
$T^{\{bb\}}_{\{\bar c\bar c\}}\to  D^-_s  \overline B^0_s l^-\bar\nu $ & $ a_4V_{cb}$& &\\
\hline
\end{tabular}
\end{table}

\subsubsection{ $\bar c\to \bar d/\bar s \ell^- \overline \nu_{\ell}$ transition}
Similarly, one can find the allowed process in hadronic level for the $\bar c\to \bar d/\bar s \ell^- \overline \nu_{\ell}$ transition. For the  channels with the $B$ meson plus $B_c$ meson in the final state, we construct the Hamiltonian as $c_1 T^{\{bb\}}_{\{\bar c\bar c\}} (H_3)_i \overline{B}^i \overline{B}_c \bar \ell \nu$.
Then the decay amplitudes are deduced as $\mathcal{M}(T^{\{bb\}}_{\{\bar c\bar c\}}\to \overline B^0  B_c^- l^-\bar\nu) = c_1 V_{\text{cd}},\; \mathcal{M}(T^{\{bb\}}_{\{\bar c\bar c\}}\to \overline B^0_s  B_c^- l^-\bar\nu) = c_1 V_{\text{cs}}$. For completeness, we give the corresponding Feynman diagram given in Fig.~\ref{fig:topology1}.(d).
%

\subsection{$T^{\{bb\}}_{\{\bar c\bar c\}}$: Non-leptonic multi-body decays }
\label{sec:nonleptonic_decay}
\begin{figure}
  \centering
  \includegraphics[width=0.75\columnwidth]{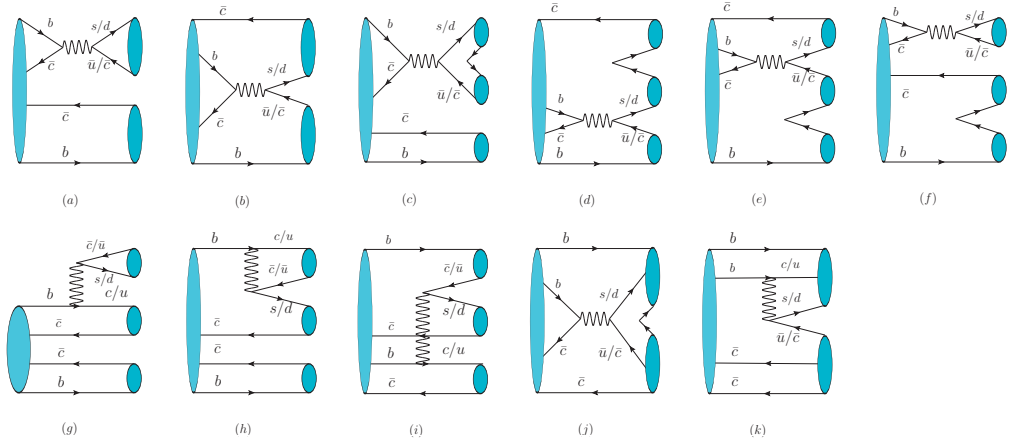}\\
  \caption{The Feynman diagrams for $T^{\{bb\}}_{\{\bar c\bar c\}}$ non-leptonic decays within $b$ quark decay. The panels(a,b) represent the two-body mesonic decays, two-body baryonic decays are shown in panels(j,k) and the panels(c-i) indicate the three-body mesonic decays.}\label{fig:topology2}
\end{figure}
\begin{figure}
  \centering
  \includegraphics[width=0.65\columnwidth]{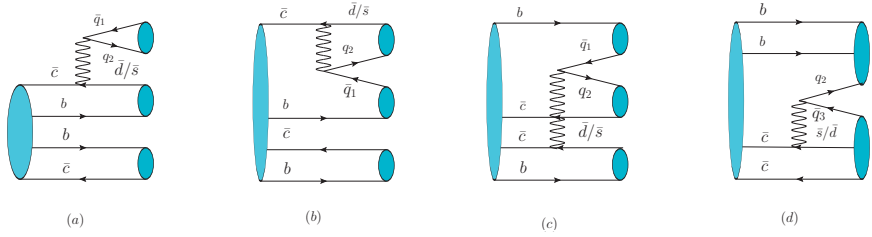}\\
  \caption{The Feynman diagrams for $T^{\{bb\}}_{\{\bar c\bar c\}}$ non-leptonic decays within $\bar c$ quark decay. The lowest allowed many-body charmed decays are three-body mesonic decays shown in panels(a,b,c) and two-body baryonic decays given in panel(d).}\label{fig:topology3}
\end{figure}

\subsubsection{$b\to c\bar c d/s$ transition}
The operators in  the transition can form an triplet under the SU(3) light quark symmetry, and  accordingly, we can write down the effective Hamiltonian of $T^{\{bb\}}_{\{\bar c\bar c\}}$ producing two or three final states as follows:
\begin{eqnarray}
&\mathcal{H}_{eff}&=a_1 T^{\{bb\}}_{\{\bar c\bar c\}} (H_3)^i D_i \overline{B}_c,\nonumber\\
&\mathcal{H}_{eff}&=a_2 T^{\{bb\}}_{\{\bar c\bar c\}} (H_3)^i D_i J/\psi \overline{B}_c+a_3 T^{\{bb\}}_{\{\bar c\bar c\}} (H_3)^i D_j M_i^j \overline{B}_c+ a_4 T^{\{bb\}}_{\{\bar c\bar c\}} (H_3)^i D_i D_j \overline{B}^j,\nonumber\\
&\mathcal{H}_{eff}&=a_5 T^{\{bb\}}_{\{\bar c\bar c\}} (H_3)^i (F_{cc})^j (\overline F_{b3})_{[ij]}+a_6 T^{\{bb\}}_{\{\bar c\bar c\}} (H_3)^i (F_{cc})^j (\overline F_{b\bar{6}})_{\{ij\}}+a_7 T^{\{bb\}}_{\{\bar c\bar c\}} (H_3)^i (\overline F_{bc})_{i} F_{ccc}.
\end{eqnarray}
The corresponding Feynman diagrams are given in  Fig.~\ref{fig:topology2}. In particular, the diagrams in  Fig.~\ref{fig:topology2}.(a,b) represent $T^{\{bb\}}_{\{\bar c\bar c\}}$ two-body mesonic decays into anti-charmed and $B_c$ mesons, and the diagrams in Fig.~\ref{fig:topology2}.(c,d) denote the three-body final states with anti-charmed meson plus $B_c$ meson and a light meson. In addition, the $a_3$ term in Hamiltonian with the  final states of two anti-charmed mesons plus $B$ meson and the $a_4$ term with the final states of a anti-charmed meson plus $B_c$ meson and $J/\psi$ are represented in several Feynman diagrams which given in Fig.~\ref{fig:topology2}.(e,f) and Fig.~\ref{fig:topology2}.(g,h,i) respectively. The two-body baryonic processes induced from $a_5,a_6,a_7$ terms are shown in Fig.~\ref{fig:topology2}.(j,k). Expanding the Hamiltonian above, one obtains the decay amplitudes which are listed in Tab.~\ref{tab:bc_Dbar_Bc23}, Tab.~\ref{tab:bc_Fccq_Fbqq}. Besides, the relations between the different decay widths are given as follows.
\begin{eqnarray*}
\Gamma(T^{\{bb\}}_{\{\bar c\bar c\}}\to \overline D^0 B_c^-\pi^- )= 6\Gamma(T^{\{bb\}}_{\{\bar c\bar c\}}\to D^- B_c^-\eta )=\Gamma(T^{\{bb\}}_{\{\bar c\bar c\}}\to  D^-_s B_c^-K^0 )=2\Gamma(T^{\{bb\}}_{\{\bar c\bar c\}}\to D^- B_c^-\pi^0 ),\\
\Gamma(T^{\{bb\}}_{\{\bar c\bar c\}}\to \overline D^0 B_c^-K^- )= \frac{3}{2}\Gamma(T^{\{bb\}}_{\{\bar c\bar c\}}\to  D^-_s B_c^-\eta )=\Gamma(T^{\{bb\}}_{\{\bar c\bar c\}}\to D^- B_c^-\overline K^0 ),\\
\Gamma(T^{\{bb\}}_{\{\bar c\bar c\}}\to B^- \overline D^0D^-)={ }\Gamma(T^{\{bb\}}_{\{\bar c\bar c\}}\to \overline B^0_s D^- D^-_s)=\frac{1}{2}\Gamma(T^{\{bb\}}_{\{\bar c\bar c\}}\to \overline B^0 D^-D^-),\\
\Gamma(T^{\{bb\}}_{\{\bar c\bar c\}}\to B^- \overline D^0 D^-_s)={ }\Gamma(T^{\{bb\}}_{\{\bar c\bar c\}}\to \overline B^0 D^- D^-_s)=\frac{1}{2}\Gamma(T^{\{bb\}}_{\{\bar c\bar c\}}\to \overline B^0_s  D^-_s D^-_s),\\
\Gamma(T^{\{bb\}}_{\{\bar c\bar c\}}\to \overline \Xi_{cc}^{--}\Sigma_{b}^{0})=
\frac{1}{2}\Gamma(T^{\{bb\}}_{\{\bar c\bar c\}}\to \overline \Xi_{cc}^{-}\Sigma_{b}^{-})=\Gamma(T^{\{bb\}}_{\{\bar c\bar c\}}\to \overline \Omega_{cc}^{-}\Xi_{b}^{\prime-}),\\
\Gamma(T^{\{bb\}}_{\{\bar c\bar c\}}\to \overline \Xi_{cc}^{--}\Xi_{b}^{\prime0})=
{ }\Gamma(T^{\{bb\}}_{\{\bar c\bar c\}}\to \overline \Xi_{cc}^{-}\Xi_{b}^{\prime-})=
\frac{1}{2}\Gamma(T^{\{bb\}}_{\{\bar c\bar c\}}\to \overline \Omega_{cc}^{-}\Omega_{b}^{-}),\\
\Gamma(T^{\{bb\}}_{\{\bar c\bar c\}}\to \overline \Xi_{cc}^{--}\Lambda_b^0)=
\Gamma(T^{\{bb\}}_{\{\bar c\bar c\}}\to \overline \Omega_{cc}^{-}\Xi_b^-),
\Gamma(T^{\{bb\}}_{\{\bar c\bar c\}}\to \overline \Xi_{cc}^{--}\Xi_b^0)= { }\Gamma(T^{\{bb\}}_{\{\bar c\bar c\}}\to \overline \Xi_{cc}^{-}\Xi_b^-).\\
\end{eqnarray*}
\begin{table}
\caption{Tetraquark $T^{\{bb\}}_{\{\bar c\bar c\}}$  decays into two and three mesons for the transition $b\to c\bar c d/s$.}\label{tab:bc_Dbar_Bc23}\begin{tabular}{|c|c|c|c|c|c|c|c}\hline\hline
channel & amplitude &channel & amplitude \\\hline
$T^{\{bb\}}_{\{\bar c\bar c\}}\to   D^-  B_c^- $ & $ a_1 V_{\text{cd}}^*$&
$T^{\{bb\}}_{\{\bar c\bar c\}}\to    D^-_s  B_c^- $ & $ a_1 V_{\text{cs}}^*$\\\hline\hline
$T^{\{bb\}}_{\{\bar c\bar c\}}\to   D^-  B_c^-  J/\psi $ & $ a_2 V_{\text{cd}}^*$&
$T^{\{bb\}}_{\{\bar c\bar c\}}\to    D^-_s  B_c^-  J/\psi $ & $ a_2 V_{\text{cs}}^*$\\\hline
\hline
$T^{\{bb\}}_{\{\bar c\bar c\}}\to   \overline D^0  D^-  B^- $ & $ a_4 V_{\text{cd}}^*$&
$T^{\{bb\}}_{\{\bar c\bar c\}}\to   \overline D^0   D^-_s  B^- $ & $ a_4 V_{\text{cs}}^*$\\\hline
$T^{\{bb\}}_{\{\bar c\bar c\}}\to   D^-  D^-  \overline B^0 $ & $ 2 a_4 V_{\text{cd}}^*$&
$T^{\{bb\}}_{\{\bar c\bar c\}}\to   D^-   D^-_s  \overline B^0 $ & $ a_4 V_{\text{cs}}^*$\\\hline
$T^{\{bb\}}_{\{\bar c\bar c\}}\to   D^-   D^-_s  \overline B^0_s $ & $ a_4 V_{\text{cd}}^*$&
$T^{\{bb\}}_{\{\bar c\bar c\}}\to    D^-_s   D^-_s  \overline B^0_s $ & $ 2 a_4 V_{\text{cs}}^*$\\\hline
\hline
$T^{\{bb\}}_{\{\bar c\bar c\}}\to   \overline D^0  B_c^-  \pi^-  $ & $ a_3 V_{\text{cd}}^*$&
$T^{\{bb\}}_{\{\bar c\bar c\}}\to   \overline D^0  B_c^-  K^-  $ & $ a_3 V_{\text{cs}}^*$\\\hline
$T^{\{bb\}}_{\{\bar c\bar c\}}\to   D^-  B_c^-  \pi^0  $ & $ -\frac{a_3 V_{\text{cd}}^*}{\sqrt{2}}$&
$T^{\{bb\}}_{\{\bar c\bar c\}}\to   D^-  B_c^-  \overline K^0  $ & $ a_3 V_{\text{cs}}^*$\\\hline
$T^{\{bb\}}_{\{\bar c\bar c\}}\to   D^-  B_c^-  \eta  $ & $ \frac{a_3 V_{\text{cd}}^*}{\sqrt{6}}$&
$T^{\{bb\}}_{\{\bar c\bar c\}}\to    D^-_s  B_c^-  K^0  $ & $ a_3 V_{\text{cd}}^*$\\\hline
$T^{\{bb\}}_{\{\bar c\bar c\}}\to    D^-_s  B_c^-  \eta  $ & $ -\sqrt{\frac{2}{3}} a_3 V_{\text{cs}}^*$& &\\
\hline \hline
\end{tabular}
\end{table}
\begin{table}
\caption{Tetraquark $T^{\{bb\}}_{\{\bar c\bar c\}}$  decays into doubly charmed baryon plus singly bottom baryon for the transition of $b\to c\bar c d/s$. }\label{tab:bc_Fccq_Fbqq}\begin{tabular}{|c|c|c|c|c|c|c|c}\hline\hline
channel & amplitude &channel & amplitude\\\hline
$T^{\{bb\}}_{\{\bar c\bar c\}}\to   \overline \Xi_{cc}^{--}  \Lambda_b^0 $ & $ -a_5 V_{\text{cd}}^*$&
$T^{\{bb\}}_{\{\bar c\bar c\}}\to   \overline \Xi_{cc}^{--}  \Xi_b^0 $ & $ -a_5 V_{\text{cs}}^*$\\\hline
$T^{\{bb\}}_{\{\bar c\bar c\}}\to   \overline \Xi_{cc}^{-}  \Xi_b^- $ & $ -a_5 V_{\text{cs}}^*$&
$T^{\{bb\}}_{\{\bar c\bar c\}}\to   \overline \Omega_{cc}^{-}  \Xi_b^- $ & $ a_5 V_{\text{cd}}^*$\\\hline
\hline
$T^{\{bb\}}_{\{\bar c\bar c\}}\to   \overline \Xi_{cc}^{--}  \Sigma_{b}^{0} $ & $ \frac{a_6 V_{\text{cd}}^*}{\sqrt{2}}$&
$T^{\{bb\}}_{\{\bar c\bar c\}}\to   \overline \Xi_{cc}^{--}  \Xi_{b}^{\prime0} $ & $ \frac{a_6 V_{\text{cs}}^*}{\sqrt{2}}$\\\hline
$T^{\{bb\}}_{\{\bar c\bar c\}}\to   \overline \Xi_{cc}^{-}  \Sigma_{b}^{-} $ & $ a_6 V_{\text{cd}}^*$&
$T^{\{bb\}}_{\{\bar c\bar c\}}\to   \overline \Xi_{cc}^{-}  \Xi_{b}^{\prime-} $ & $ \frac{a_6 V_{\text{cs}}^*}{\sqrt{2}}$\\\hline
$T^{\{bb\}}_{\{\bar c\bar c\}}\to   \overline \Omega_{cc}^{-}  \Xi_{b}^{\prime-} $ & $ \frac{a_6 V_{\text{cd}}^*}{\sqrt{2}}$&
$T^{\{bb\}}_{\{\bar c\bar c\}}\to   \overline \Omega_{cc}^{-}  \Omega_{b}^{-} $ & $ a_6 V_{\text{cs}}^*$\\\hline
\hline
$T^{\{bb\}}_{\{\bar c\bar c\}}\to   \Xi_{bc}^{0}  \overline \Omega_{ccc}^{--} $ & $ a_7 V_{\text{cd}}^*$&
$T^{\{bb\}}_{\{\bar c\bar c\}}\to   \Omega_{bc}^{0}  \overline \Omega_{ccc}^{--} $ & $ a_7 V_{\text{cs}}^*$\\\hline
\hline
\end{tabular}
\end{table}
\subsubsection{$b\to c \bar u d/s$ transition}
The hadron-level effective Hamiltonian of two-body and three-body decays can be constructed as
\begin{eqnarray}
&\mathcal{H}_{eff}&=b_1T^{\{bb\}}_{\{\bar c\bar c\}} (H_8)^i_j M^j_i \overline{B}_c + b_2 T^{\{bb\}}_{\{\bar c\bar c\}} (H_8)^i_j \overline{B}^j D_i ,\nonumber\\
&\mathcal{H}_{eff}&=b_3T^{\{bb\}}_{\{\bar c\bar c\}} (H_8)^i_j M^j_i J/\psi \overline{B}_c + b_4 T^{\{bb\}}_{\{\bar c\bar c\}} (H_8)^i_j \overline{D}^j D_i \overline{B}_c +b_5T^{\{bb\}}_{\{\bar c\bar c\}} (H_8)^i_j \overline{B}^j D_i J/\psi \nonumber\\
&&+b_6 T^{\{bb\}}_{\{\bar c\bar c\}} (H_8)^i_j \overline{B}^j D_k M^k_i+b_7 T^{\{bb\}}_{\{\bar c\bar c\}} (H_8)^i_k \overline{B}^j D_j M^k_i+b_8 T^{\{bb\}}_{\{\bar c\bar c\}} (H_8)^i_k \overline{B}^j D_i M^k_j\nonumber\\
&&+b_9 T^{\{bb\}}_{\{\bar c\bar c\}} (H_8)^i_j \overline{B}_c M^k_i M^j_k,\nonumber\\
&\mathcal{H}_{eff}&=b_{10}T^{\{bb\}}_{\{\bar c\bar c\}} (H_8)^i_j (F_{c\bar3})^{[jk]} (\overline F_{b3})_{[ik]}+b_{11}T^{\{bb\}}_{\{\bar c\bar c\}} (H_8)^i_j (F_{c\bar3})^{[jk]} (\overline F_{b\bar{6}})_{\{ik\}}\nonumber\\
&&+b_{12}T^{\{bb\}}_{\{\bar c\bar c\}} (H_8)^i_j (F_{c6})^{\{jk\}} (\overline F_{b3})_{[ik]}+b_{13}T^{\{bb\}}_{\{\bar c\bar c\}} (H_8)^i_j (F_{c6})^{\{jk\}} (\overline F_{b\bar{6}})_{\{ik\}}\nonumber\\
&&+b_{14}T^{\{bb\}}_{\{\bar c\bar c\}} (H_8)^i_j (F_{cc})^{j} (\overline F_{bc})_{i}.
\end{eqnarray}
At the topological level, the relevant Feynman diagrams are shown in Fig.~\ref{fig:topology2}. One derives the decay amplitudes given in Tab.~\ref{tab:bc_Bc_M}, Tab.~\ref{tab:bc_Fcqq_Fbqq} respectively. Accordingly, we obtain the relations between different decay widths as follows:
\begin{eqnarray*}
\Gamma(T^{\{bb\}}_{\{\bar c\bar c\}}\to B_c^- \pi^0 K^- )= 3\Gamma(T^{\{bb\}}_{\{\bar c\bar c\}}\to B_c^- K^- \eta )=
\frac{1}{2}\Gamma(T^{\{bb\}}_{\{\bar c\bar c\}}\to B_c^- \pi^- \overline K^0 ),\\
\Gamma(T^{\{bb\}}_{\{\bar c\bar c\}}\to B_c^- K^0 K^- )= \frac{3}{2}\Gamma(T^{\{bb\}}_{\{\bar c\bar c\}}\to B_c^- \pi^- \eta ),
\Gamma(T^{\{bb\}}_{\{\bar c\bar c\}}\to B^-  D^-_s\pi^0 )= \frac{1}{2}\Gamma(T^{\{bb\}}_{\{\bar c\bar c\}}\to \overline B^0  D^-_s\pi^- ),\\
\Gamma(T^{\{bb\}}_{\{\bar c\bar c\}}\to \overline \Sigma_{c}^{--}\Sigma_{b}^{0})= { }\Gamma(T^{\{bb\}}_{\{\bar c\bar c\}}\to \overline \Sigma_{c}^{-}\Sigma_{b}^{-})=2\Gamma(T^{\{bb\}}_{\{\bar c\bar c\}}\to \overline \Xi_{c}^{\prime-}\Xi_{b}^{\prime-}),\\
\Gamma(T^{\{bb\}}_{\{\bar c\bar c\}}\to \overline \Sigma_{c}^{--}\Xi_{b}^{\prime0})=2\Gamma(T^{\{bb\}}_{\{\bar c\bar c\}}\to \overline \Sigma_{c}^{-}\Xi_{b}^{\prime-})=\Gamma(T^{\{bb\}}_{\{\bar c\bar c\}}\to \overline \Xi_{c}^{\prime-}\Omega_{b}^{-}),\\
\Gamma(T^{\{bb\}}_{\{\bar c\bar c\}}\to \overline \Lambda_{c}^-\Sigma_{b}^{-})=2\Gamma(T^{\{bb\}}_{\{\bar c\bar c\}}\to \overline \Xi_{c}^-\Xi_{b}^{\prime-}),
\Gamma(T^{\{bb\}}_{\{\bar c\bar c\}}\to \overline \Lambda_{c}^-\Xi_{b}^{\prime-})=\frac{1}{2}\Gamma(T^{\{bb\}}_{\{\bar c\bar c\}}\to \overline \Xi_{c}^-\Omega_{b}^{-}),\\
\Gamma(T^{\{bb\}}_{\{\bar c\bar c\}}\to \overline \Sigma_{c}^{--}\Lambda_b^0)=2\Gamma(T^{\{bb\}}_{\{\bar c\bar c\}}\to \overline \Xi_{c}^{\prime-}\Xi_b^-),
\Gamma(T^{\{bb\}}_{\{\bar c\bar c\}}\to \overline \Sigma_{c}^{--}\Xi_b^0)=2\Gamma(T^{\{bb\}}_{\{\bar c\bar c\}}\to \overline \Sigma_{c}^{-}\Xi_b^-).\\
\end{eqnarray*}
\begin{table}
\caption{Tetraquark $T^{\{bb\}}_{\{\bar c\bar c\}}$ decays into two mesons and three mesons for the transition $b\to c \bar u d/s$.}\label{tab:bc_Bc_M}\begin{tabular}{|c|c|c|c|c|c|c|c}\hline\hline
channel & amplitude&channel & amplitude \\\hline
$T^{\{bb\}}_{\{\bar c\bar c\}}\to   \pi^-   B_c^- $ & $ b_1 V_{\text{ud}}^*$&
$T^{\{bb\}}_{\{\bar c\bar c\}}\to   K^-   B_c^- $ & $ b_1 V_{\text{us}}^*$\\\hline
$T^{\{bb\}}_{\{\bar c\bar c\}}\to   B^-  D^- $ & $ b_2 V_{\text{ud}}^*$&
$T^{\{bb\}}_{\{\bar c\bar c\}}\to   B^-   D^-_s $ & $ b_2 V_{\text{us}}^*$\\\hline
\hline
$T^{\{bb\}}_{\{\bar c\bar c\}}\to   \pi^-   J/\psi  B_c^- $ & $ b_3 V_{\text{ud}}^*$&
$T^{\{bb\}}_{\{\bar c\bar c\}}\to   K^-   J/\psi  B_c^- $ & $ b_3 V_{\text{us}}^*$\\\hline
$T^{\{bb\}}_{\{\bar c\bar c\}}\to    D^0  D^-  B_c^- $ & $ b_4 V_{\text{ud}}^*$&
$T^{\{bb\}}_{\{\bar c\bar c\}}\to    D^0   D^-_s  B_c^- $ & $ b_4 V_{\text{us}}^*$\\\hline
$T^{\{bb\}}_{\{\bar c\bar c\}}\to   B^-  D^-  J/\psi $ & $ b_5 V_{\text{ud}}^*$&
$T^{\{bb\}}_{\{\bar c\bar c\}}\to   B^-   D^-_s  J/\psi $ & $ b_5 V_{\text{us}}^*$\\\hline
\hline
$T^{\{bb\}}_{\{\bar c\bar c\}}\to   B^-  \overline D^0  \pi^-  $ & $ \left(b_6+b_7\right) V_{\text{ud}}^*$&
$T^{\{bb\}}_{\{\bar c\bar c\}}\to   B^-  \overline D^0  K^-  $ & $ \left(b_6+b_7\right) V_{\text{us}}^*$\\\hline
$T^{\{bb\}}_{\{\bar c\bar c\}}\to   B^-  D^-  \pi^0  $ & $ \frac{\left(b_8-b_6\right) V_{\text{ud}}^*}{\sqrt{2}}$&
$T^{\{bb\}}_{\{\bar c\bar c\}}\to   B^-  D^-  \overline K^0  $ & $ b_6 V_{\text{us}}^*$\\\hline
$T^{\{bb\}}_{\{\bar c\bar c\}}\to   B^-  D^-  \eta  $ & $ \frac{\left(b_6+b_8\right) V_{\text{ud}}^*}{\sqrt{6}}$&
$T^{\{bb\}}_{\{\bar c\bar c\}}\to   B^-   D^-_s  \pi^0  $ & $ \frac{b_8 V_{\text{us}}^*}{\sqrt{2}}$\\\hline
$T^{\{bb\}}_{\{\bar c\bar c\}}\to   B^-   D^-_s  K^0  $ & $ b_6 V_{\text{ud}}^*$&
$T^{\{bb\}}_{\{\bar c\bar c\}}\to   B^-   D^-_s  \eta  $ & $ \frac{\left(b_8-2 b_6\right) V_{\text{us}}^*}{\sqrt{6}}$\\\hline
$T^{\{bb\}}_{\{\bar c\bar c\}}\to   \overline B^0  D^-  \pi^-  $ & $ \left(b_7+b_8\right) V_{\text{ud}}^*$&
$T^{\{bb\}}_{\{\bar c\bar c\}}\to   \overline B^0  D^-  K^-  $ & $ b_7 V_{\text{us}}^*$\\\hline
$T^{\{bb\}}_{\{\bar c\bar c\}}\to   \overline B^0   D^-_s  \pi^-  $ & $ b_8 V_{\text{us}}^*$&
$T^{\{bb\}}_{\{\bar c\bar c\}}\to   \overline B^0_s  D^-  K^-  $ & $ b_8 V_{\text{ud}}^*$\\\hline
$T^{\{bb\}}_{\{\bar c\bar c\}}\to   \overline B^0_s   D^-_s  \pi^-  $ & $ b_7 V_{\text{ud}}^*$&
$T^{\{bb\}}_{\{\bar c\bar c\}}\to   \overline B^0_s   D^-_s  K^-  $ & $ \left(b_7+b_8\right) V_{\text{us}}^*$\\\hline
\hline
$T^{\{bb\}}_{\{\bar c\bar c\}}\to   \pi^0   K^-   B_c^- $ & $ \frac{b_9 V_{\text{us}}^*}{\sqrt{2}}$&
$T^{\{bb\}}_{\{\bar c\bar c\}}\to   \pi^-   \overline K^0   B_c^- $ & $ b_9 V_{\text{us}}^*$\\\hline
$T^{\{bb\}}_{\{\bar c\bar c\}}\to   \pi^-   \eta   B_c^- $ & $ \sqrt{\frac{2}{3}} b_9 V_{\text{ud}}^*$&
$T^{\{bb\}}_{\{\bar c\bar c\}}\to   K^0   K^-   B_c^- $ & $ b_9 V_{\text{ud}}^*$\\\hline
$T^{\{bb\}}_{\{\bar c\bar c\}}\to   K^-   \eta   B_c^- $ & $ -\frac{b_9 V_{\text{us}}^*}{\sqrt{6}}$& &\\
\hline
\end{tabular}
\end{table}
\begin{table}
\caption{Tetraquark $T^{\{bb\}}_{\{\bar c\bar c\}}$ decays into singly charmed baryon plus singly bottom baryon for the transition $b\to c \bar u d/s$.}\label{tab:bc_Fcqq_Fbqq}\begin{tabular}{|c|c|c|c|c|c|c|c}\hline\hline
channel & amplitude&channel & amplitude \\\hline
$T^{\{bb\}}_{\{\bar c\bar c\}}\to   \overline \Lambda_{c}^-  \Xi_b^- $ & $ -b_{10} V_{\text{us}}^*$&
$T^{\{bb\}}_{\{\bar c\bar c\}}\to   \overline \Xi_{c}^-  \Xi_b^- $ & $ b_{10} V_{\text{ud}}^*$\\\hline
\hline
$T^{\{bb\}}_{\{\bar c\bar c\}}\to   \overline \Lambda_{c}^-  \Sigma_{b}^{-} $ & $ b_{11} V_{\text{ud}}^*$&
$T^{\{bb\}}_{\{\bar c\bar c\}}\to   \overline \Lambda_{c}^-  \Xi_{b}^{\prime-} $ & $ \frac{b_{11} V_{\text{us}}^*}{\sqrt{2}}$\\\hline
$T^{\{bb\}}_{\{\bar c\bar c\}}\to   \overline \Xi_{c}^-  \Xi_{b}^{\prime-} $ & $ \frac{b_{11} V_{\text{ud}}^*}{\sqrt{2}}$&
$T^{\{bb\}}_{\{\bar c\bar c\}}\to   \overline \Xi_{c}^-  \Omega_{b}^{-} $ & $ b_{11} V_{\text{us}}^*$\\\hline
\hline$T^{\{bb\}}_{\{\bar c\bar c\}}\to   \overline \Sigma_{c}^{--}  \Lambda_b^0 $ & $ -b_{12} V_{\text{ud}}^*$&
$T^{\{bb\}}_{\{\bar c\bar c\}}\to   \overline \Sigma_{c}^{--}  \Xi_b^0 $ & $ -b_{12} V_{\text{us}}^*$\\\hline
$T^{\{bb\}}_{\{\bar c\bar c\}}\to   \overline \Sigma_{c}^{-}  \Xi_b^- $ & $ -\frac{b_{12} V_{\text{us}}^*}{\sqrt{2}}$&
$T^{\{bb\}}_{\{\bar c\bar c\}}\to   \overline \Xi_{c}^{\prime-}  \Xi_b^- $ & $ \frac{b_{12} V_{\text{ud}}^*}{\sqrt{2}}$\\\hline
\hline$T^{\{bb\}}_{\{\bar c\bar c\}}\to   \overline \Sigma_{c}^{--}  \Sigma_{b}^{0} $ & $ \frac{b_{13} V_{\text{ud}}^*}{\sqrt{2}}$&
$T^{\{bb\}}_{\{\bar c\bar c\}}\to   \overline \Sigma_{c}^{--}  \Xi_{b}^{\prime0} $ & $ \frac{b_{13} V_{\text{us}}^*}{\sqrt{2}}$\\\hline
$T^{\{bb\}}_{\{\bar c\bar c\}}\to   \overline \Sigma_{c}^{-}  \Sigma_{b}^{-} $ & $ \frac{b_{13} V_{\text{ud}}^*}{\sqrt{2}}$&
$T^{\{bb\}}_{\{\bar c\bar c\}}\to   \overline \Sigma_{c}^{-}  \Xi_{b}^{\prime-} $ & $ \frac{1}{2} b_{13} V_{\text{us}}^*$\\\hline
$T^{\{bb\}}_{\{\bar c\bar c\}}\to   \overline \Xi_{c}^{\prime-}  \Xi_{b}^{\prime-} $ & $ \frac{1}{2} b_{13} V_{\text{ud}}^*$&
$T^{\{bb\}}_{\{\bar c\bar c\}}\to   \overline \Xi_{c}^{\prime-}  \Omega_{b}^{-} $ & $ \frac{b_{13} V_{\text{us}}^*}{\sqrt{2}}$\\\hline
\hline
$T^{\{bb\}}_{\{\bar c\bar c\}}\to   \overline \Xi_{cc}^{--}  \Xi_{bc}^{0} $ & $ b_{14} V_{\text{ud}}^*$&
$T^{\{bb\}}_{\{\bar c\bar c\}}\to   \overline \Xi_{cc}^{--}  \Omega_{bc}^{0} $ & $ b_{14} V_{\text{us}}^*$\\\hline
\hline
\end{tabular}
\end{table}
\subsubsection{ $b\to u \bar c d/s$ transition}
The effective Hamiltonian at the hadron level for $T^{\{bb\}}_{\{\bar c\bar c\}}$ producing three mesons or two baryons are constructed as
\begin{eqnarray}
&\mathcal{H}_{eff}&=c_1 T^{\{bb\}}_{\{\bar c\bar c\}} (H_{6} )^{\{ij\}}D_i D_j \overline{B}_c,\nonumber\\
&\mathcal{H}_{eff}&=c_2 T^{\{bb\}}_{\{\bar c\bar c\}} (H_{\bar 3} )^{[ij]} (\overline F_{b3})_{[ij]} F_{ccc}+c_3 T^{\{bb\}}_{\{\bar c\bar c\}} (H_{6} )^{\{ij\}} (\overline F_{b\bar{6}})_{\{ij\}} F_{ccc}.
\end{eqnarray}
It should be noticed that the operator $H_{\bar3}$ in mesonic process vanishs as the two antisymmetry superscripts contract with the two symmetry anti-charmed fields. Though the Hamiltonian for the mesonic process follows only $c_1$ term, the corresponding Feynman diagrams can be allowed with different topologies given in Fig.~\ref{fig:topology2}.(g,h,i). One then proceed to obtain the decay amplitudes $\mathcal{M}(T^{\{bb\}}_{\{\bar c\bar c\}}\to   \overline D^0  D^-  B_c^- )=2 c_1 V_{\text{cd}}^*$,
$\mathcal{M}(T^{\{bb\}}_{\{\bar c\bar c\}}\to   \overline D^0   D^-_s  B_c^- )=2 c_1 V_{\text{cs}}^*$ for the mesonic processes and $\mathcal{M}(T^{\{bb\}}_{\{\bar c\bar c\}}\to   \Lambda_b^0  \overline \Omega_{ccc}^{--}) = 2 c_2 V_{\text{cd}}^*$, $\mathcal{M}(T^{\{bb\}}_{\{\bar c\bar c\}}\to   \Xi_b^0  \overline \Omega_{ccc}^{--})= 2 c_2 V_{\text{cs}}^*$, $\mathcal{M}(T^{\{bb\}}_{\{\bar c\bar c\}}\to   \Sigma_{b}^{0}  \overline \Omega_{ccc}^{--})= \sqrt{2} c_3 V_{\text{cd}}^*$, $\mathcal{M}(T^{\{bb\}}_{\{\bar c\bar c\}}\to   \Xi_{b}^{\prime0}  \overline \Omega_{ccc}^{--})=\sqrt{2} c_3 V_{\text{cs}}^*$ for the baryonic processes, from which we derive the equation as
\begin{eqnarray*}
\frac{\Gamma(T^{\{bb\}}_{\{\bar c\bar c\}}\to   \overline D^0  D^-  B_c^- )}{\Gamma(T^{\{bb\}}_{\{\bar c\bar c\}}\to   \overline D^0   D^-_s  B_c^- )}=\frac{\Gamma(T^{\{bb\}}_{\{\bar c\bar c\}}\to   \Lambda_b^0  \overline \Omega_{ccc}^{--} )}{\Gamma(T^{\{bb\}}_{\{\bar c\bar c\}}\to   \Xi_b^0  \overline \Omega_{ccc}^{--})}=\frac{\Gamma(T^{\{bb\}}_{\{\bar c\bar c\}}\to   \Sigma_{b}^{0}  \overline \Omega_{ccc}^{--})}{\Gamma(T^{\{bb\}}_{\{\bar c\bar c\}}\to   \Xi_{b}^{\prime0}  \overline \Omega_{ccc}^{--})}=\frac{|V_{cd}|^2}{|V_{cs}|^2}.
\end{eqnarray*}

\subsubsection{$b\to q_1 \bar q_2 q_3$ Charmless transition}
At the hadron  level, the effective Hamiltonian for $T^{\{bb\}}_{\{\bar c\bar c\}}$ decaying into mesons or baryons is constructed  as follows,
\begin{eqnarray}
&\mathcal{H}_{eff}&=d_1 T^{\{bb\}}_{\{\bar c\bar c\}} (H_3 )^{i} \overline{B}^j D_i D_j +d_2 T^{\{bb\}}_{\{\bar c\bar c\}} (H_{15} )^{\{ij\}}_k \overline{B}^k D_i D_j,\nonumber\\
&\mathcal{H}_{eff}&=d_3 T^{\{bb\}}_{\{\bar c\bar c\}} (H_3 )^{i} M^j_i D_j \overline{B}_c+d_4 T^{\{bb\}}_{\{\bar c\bar c\}}(H_{\bar6} )^{[ij]}_k M^k_i D_j \overline{B}_c+d_5 T^{\{bb\}}_{\{\bar c\bar c\}} (H_{15} )^{\{ij\}}_k M^k_i D_j \overline{B}_c,\nonumber\\
&\mathcal{H}_{eff}&=d_6 T^{\{bb\}}_{\{\bar c\bar c\}} (H_3 )^{i} (F_{cc})^j (\overline F_{b3})_{[ij]}+d_7 T^{\{bb\}}_{\{\bar c\bar c\}} (H_{\bar6} )^{[ij]}_{k} (F_{cc})^k (\overline F_{b3})_{[ij]}\nonumber\\
&&+d_8 T^{\{bb\}}_{\{\bar c\bar c\}} (H_3 )^{i} (F_{cc})^j (\overline F_{b\bar{6}})_{\{ij\}}+d_9 T^{\{bb\}}_{\{\bar c\bar c\}} (H_{15} )^{\{ij\}}_{k} (F_{cc})^k (\overline F_{b\bar{6}})_{\{ij\}}.
\end{eqnarray}
In the three-body mesonic decays, the decay amplitudes are given in Tab.~\ref{tab:bc_Dbar_D_B_d} for the transition $b\to d$ and Tab.~\ref{tab:bc_Dbar_D_B_s} for the transition $b\to s$.  In the two-body baryonic decays, the corresponding amplitudes are listed in  Tab.~\ref{tab:bc_Fccq_Fbqq_d} for the transition $b\to d$ and Tab.~\ref{tab:bc_Fccq_Fbqq_s} for the transition $b\to s$. We obtain the relations of these decay widths given as
\begin{eqnarray*}
\Gamma(T^{\{bb\}}_{\{\bar c\bar c\}}\to \overline B^0 D^-D^-)= 2\Gamma(T^{\{bb\}}_{\{\bar c\bar c\}}\to \overline B^0_s D^- D^-_s),
\Gamma(T^{\{bb\}}_{\{\bar c\bar c\}}\to \overline B^0 D^- D^-_s)= \frac{1}{2}\Gamma(T^{\{bb\}}_{\{\bar c\bar c\}}\to \overline B^0_s  D^-_s D^-_s),\\
\Gamma(T^{\{bb\}}_{\{\bar c\bar c\}}\to \overline \Xi_{cc}^{-}\Sigma_{b}^{-})= 2\Gamma(T^{\{bb\}}_{\{\bar c\bar c\}}\to \overline \Omega_{cc}^{-}\Xi_{b}^{\prime-}),
\Gamma(T^{\{bb\}}_{\{\bar c\bar c\}}\to \overline \Xi_{cc}^{-}\Xi_{b}^{\prime-})= \frac{1}{2}\Gamma(T^{\{bb\}}_{\{\bar c\bar c\}}\to \overline \Omega_{cc}^{-}\Omega_{b}^{-}).
\end{eqnarray*}
\begin{table}
\caption{Tetraquark $T^{\{bb\}}_{\{\bar c\bar c\}}$ decays into three mesons induced by the charmless $b\to d$ transition.}\label{tab:bc_Dbar_D_B_d}\begin{tabular}{|c|c|c|c|c|c|c|c}\hline\hline
channel & amplitude&channel & amplitude \\\hline
$T^{\{bb\}}_{\{\bar c\bar c\}}\to   \overline D^0  \pi^-   B_c^- $ & $ d_3-d_4+3 d_5$&
$T^{\{bb\}}_{\{\bar c\bar c\}}\to   D^-  \pi^0   B_c^- $ & $ \frac{-d_3+d_4+5 d_5}{\sqrt{2}}$\\\hline
$T^{\{bb\}}_{\{\bar c\bar c\}}\to   D^-  \eta   B_c^- $ & $ \frac{d_3+3 \left(d_4+d_5\right)}{\sqrt{6}}$&
$T^{\{bb\}}_{\{\bar c\bar c\}}\to    D^-_s  K^0   B_c^- $ & $ d_3+d_4-d_5$\\\hline
\hline
$T^{\{bb\}}_{\{\bar c\bar c\}}\to   \overline D^0  D^-  B^- $ & $  \left(d_1+6 d_2\right)$&
$T^{\{bb\}}_{\{\bar c\bar c\}}\to   D^-  D^-  \overline B^0 $ & $ 2 \left(d_1-2 d_2\right)$\\\hline
$T^{\{bb\}}_{\{\bar c\bar c\}}\to   D^-   D^-_s  \overline B^0_s $ & $ \left(d_1-2 d_2\right)$& &\\
\hline\hline
\end{tabular}
\end{table}
\begin{table}
\caption{Tetraquark $T^{\{bb\}}_{\{\bar c\bar c\}}$ decays into three mesons induced by the charmless $b\to s$ transition.}\label{tab:bc_Dbar_D_B_s}\begin{tabular}{|c|c|c|c|c|c|c|c}\hline\hline
channel & amplitude&channel & amplitude \\\hline
$T^{\{bb\}}_{\{\bar c\bar c\}}\to   \overline D^0  K^-   B_c^- $ & $ d_3-d_4+3 d_5$&
$T^{\{bb\}}_{\{\bar c\bar c\}}\to   D^-  \overline K^0   B_c^- $ & $ d_3+d_4-d_5$\\\hline
$T^{\{bb\}}_{\{\bar c\bar c\}}\to    D^-_s  \pi^0   B_c^- $ & $ \sqrt{2} \left(d_4+2 d_5\right)$&
$T^{\{bb\}}_{\{\bar c\bar c\}}\to    D^-_s  \eta   B_c^- $ & $ -\sqrt{\frac{2}{3}} \left(d_3-3 d_5\right)$\\\hline
\hline
$T^{\{bb\}}_{\{\bar c\bar c\}}\to   \overline D^0   D^-_s  B^- $ & $  \left(d_1+6 d_2\right)$&
$T^{\{bb\}}_{\{\bar c\bar c\}}\to   D^-   D^-_s  \overline B^0 $ & $  \left(d_1-2 d_2\right)$\\\hline
$T^{\{bb\}}_{\{\bar c\bar c\}}\to    D^-_s   D^-_s  \overline B^0_s $ & $2 \left(d_1-2 d_2\right)$& &\\
\hline\hline
\end{tabular}
\end{table}
\begin{table}
\caption{Tetraquark $T^{\{bb\}}_{\{\bar c\bar c\}}$ decays into doubly charmed baryon plus singly bottom baryon induced by the charmless $b\to d$ transition .}\label{tab:bc_Fccq_Fbqq_d}\begin{tabular}{|c|c|c|c|c|c|c|c}\hline\hline
channel & amplitude &channel & amplitude\\\hline
$T^{\{bb\}}_{\{\bar c\bar c\}}\to   \overline \Xi_{cc}^{--}  \Lambda_b^0 $ & $ 2 d_7-d_6$&
$T^{\{bb\}}_{\{\bar c\bar c\}}\to   \overline \Omega_{cc}^{-}  \Xi_b^- $ & $ d_6+2 d_7$\\\hline
\hline
$T^{\{bb\}}_{\{\bar c\bar c\}}\to   \overline \Xi_{cc}^{--}  \Sigma_{b}^{0} $ & $ \frac{d_8+6 d_9}{\sqrt{2}}$&
$T^{\{bb\}}_{\{\bar c\bar c\}}\to   \overline \Xi_{cc}^{-}  \Sigma_{b}^{-} $ & $ d_8-2 d_9$\\\hline
$T^{\{bb\}}_{\{\bar c\bar c\}}\to   \overline \Omega_{cc}^{-}  \Xi_{b}^{\prime-} $ & $ \frac{d_8-2 d_9}{\sqrt{2}}$& &\\\hline
\hline
\end{tabular}
\end{table}
\begin{table}
\caption{Tetraquark $T^{\{bb\}}_{\{\bar c\bar c\}}$ decays into doubly charmed baryon plus singly bottom baryon induced by the charmless $b\to s$ transition .}\label{tab:bc_Fccq_Fbqq_s}\begin{tabular}{|c|c|c|c|c|c|c|c}\hline\hline
channel & amplitude&channel & amplitude \\\hline
$T^{\{bb\}}_{\{\bar c\bar c\}}\to   \overline \Xi_{cc}^{--}  \Xi_b^0 $ & $ 2 d_7-d_6$&
$T^{\{bb\}}_{\{\bar c\bar c\}}\to   \overline \Xi_{cc}^{-}  \Xi_b^- $ & $ -d_6-2 d_7$\\\hline
\hline
$T^{\{bb\}}_{\{\bar c\bar c\}}\to   \overline \Xi_{cc}^{--}  \Xi_{b}^{\prime0} $ & $ \frac{d_8+6 d_9}{\sqrt{2}}$&
$T^{\{bb\}}_{\{\bar c\bar c\}}\to   \overline \Xi_{cc}^{-}  \Xi_{b}^{\prime-} $ & $ \frac{d_8-2d_9}{\sqrt{2}}$\\\hline
$T^{\{bb\}}_{\{\bar c\bar c\}}\to   \overline \Omega_{cc}^{-}  \Omega_{b}^{-} $ & $ d_8-2 d_9$ &&\\\hline
\hline
\end{tabular}
\end{table}
\subsubsection{$\bar c\to \bar q_1 q_2 \bar q_3$ transition}
The effective Hamiltonian at the hadron-level for $T_{\{\bar c\bar c\}}^{\{bb\}}$ producing two or three body final states can be constructed as follows,
\begin{eqnarray}
&\mathcal{H}_{eff}&=f_1 T^{\{bb\}}_{\{\bar c\bar c\}} (H_{\overline {15}})^k_{\{ij\}} \overline B^i \overline B^j D_k,\nonumber\\
&\mathcal{H}_{eff}&=f_2 T^{\{bb\}}_{\{\bar c\bar c\}} (H_{6})^k_{[ij]} M^i_k \overline{B}^j \overline{B}_c +f_3 T^{\{bb\}}_{\{\bar c\bar c\}} (H_{\overline {15}})^k_{\{ij\}} M^i_k \overline B^j \overline{B}_c,\nonumber\\
&\mathcal{H}_{eff}&=f_4 T^{\{bb\}}_{\{\bar c\bar c\}} (H_{6})^k_{[ij]} (F_{c\bar3})^{[ij]} (\overline F_{bb})_{k} +f_5 T^{\{bb\}}_{\{\bar c\bar c\}} (H_{\overline{15}})^k_{\{ij\}} (F_{c6})^{\{ij\}} (\overline F_{bb})_{k}.
\end{eqnarray}
Here, it should be noticed that the above effective Hamiltonian can not lead to the two-body mesonic decays of $T^{\{bb\}}_{\{\bar c\bar c\}}$. Further more, the corresponding Feynman diagrams are given in Fig.~\ref{fig:topology3}. Expanding the Hamiltonian above and we can obtain the decay amplitudes shown in Tab.~\ref{tab:bc_2B_Dbar}, and Tab.~\ref{tab:bc_Fcqq_Fbbq_c}. The relations between different decay widths are given as
\begin{eqnarray*}
\Gamma(T^{\{bb\}}_{\{\bar c\bar c\}}\to D^- B^-\overline B^0)= { }\Gamma(T^{\{bb\}}_{\{\bar c\bar c\}}\to  D^-_s B^-\overline B^0_s),
\Gamma(T^{\{bb\}}_{\{\bar c\bar c\}}\to B^- \pi^0 B_c^-)= \frac{1}{3}\Gamma(T^{\{bb\}}_{\{\bar c\bar c\}}\to B^- \eta B_c^-),\\
\Gamma(T^{\{bb\}}_{\{\bar c\bar c\}}\to \overline B^0 \pi^- B_c^-)= \Gamma(T^{\{bb\}}_{\{\bar c\bar c\}}\to \overline B^0_s K^- B_c^-),\\
\Gamma(T^{\{bb\}}_{\{\bar c\bar c\}}\to  \overline \Lambda_{c}^-  \Xi_{bb}^{-}) = \Gamma(T^{\{bb\}}_{\{\bar c\bar c\}}\to   \overline \Xi_{c}^-  \Omega_{bb}^{-}),
\Gamma(T^{\{bb\}}_{\{\bar c\bar c\}}\to   \overline \Sigma_{c}^{-}  \Xi_{bb}^{-})= \Gamma(T^{\{bb\}}_{\{\bar c\bar c\}}\to   \overline \Xi_{c}^{\prime-}  \Omega_{bb}^{-}).
\end{eqnarray*}
\begin{table}
\caption{Tetraquark $T^{\{bb\}}_{\{\bar c\bar c\}}$  decays into three mesons for the transition $\bar c\to \bar q_1 q_2 \bar q_3$. In particular, the amplitudes are shown as Cabibbo allowed, singly Cabibbo suppressed, doubly suppressed respectively. }\label{tab:bc_2B_Dbar}\begin{tabular}{|c|c|c|c|c|c|c|c}\hline\hline
channel & amplitude &channel & amplitude \\\hline
$T^{\{bb\}}_{\{\bar c\bar c\}}\to   B^-  \overline B^0_s  D^- $ & $ 2 f_1$&
$T^{\{bb\}}_{\{\bar c\bar c\}}\to   B^-  K^0   B_c^- $ & $ f_2+f_3$\\\hline
$T^{\{bb\}}_{\{\bar c\bar c\}}\to   \overline B^0_s  \pi^-   B_c^- $ & $ f_3-f_2$& &\\
\hline\hline
$T^{\{bb\}}_{\{\bar c\bar c\}}\to   B^-  \overline B^0  D^- $ & $ -2 f_1 \text{sC}$&
$T^{\{bb\}}_{\{\bar c\bar c\}}\to   B^-  \overline B^0_s   D^-_s $ & $ 2 f_1 \text{sC}$\\\hline
$T^{\{bb\}}_{\{\bar c\bar c\}}\to   B^-  \pi^0   B_c^- $ & $ \frac{\left(f_2+f_3\right) \text{sC}}{\sqrt{2}}$&
$T^{\{bb\}}_{\{\bar c\bar c\}}\to   B^-  \eta   B_c^- $ & $ -\sqrt{\frac{3}{2}} \left(f_2+f_3\right) \text{sC}$\\\hline
$T^{\{bb\}}_{\{\bar c\bar c\}}\to   \overline B^0  \pi^-   B_c^- $ & $ \left(f_2-f_3\right) \text{sC}$&
$T^{\{bb\}}_{\{\bar c\bar c\}}\to   \overline B^0_s  K^-   B_c^- $ & $ \left(f_3-f_2\right) \text{sC}$\\\hline
\hline
$T^{\{bb\}}_{\{\bar c\bar c\}}\to   B^-  \overline B^0   D^-_s $ & $ 2 f_1 \text{sC}^2$&
$T^{\{bb\}}_{\{\bar c\bar c\}}\to   B^-  \overline K^0   B_c^- $ & $ \left(f_2+f_3\right) \text{sC}^2$\\\hline
$T^{\{bb\}}_{\{\bar c\bar c\}}\to   \overline B^0  K^-   B_c^- $ & $ \left(f_2-f_3\right) \left(-\text{sC}^2\right)$& &\\
\hline
\end{tabular}
\end{table}
\begin{table}
\caption{Tetraquark $T^{\{bb\}}_{\{\bar c\bar c\}}$  decays into singly charmed baryon and doubly bottom baryon for the transition $\bar c\to\bar q_1 q_2\bar q_3$.  In particular, the amplitudes are shown as Cabibbo allowed, singly Cabibbo suppressed, doubly suppressed respectively.}\label{tab:bc_Fcqq_Fbbq_c}\begin{tabular}{|c|c|c|c|c|c|c|c}\hline\hline
channel & amplitude&channel & amplitude  \\\hline
$T^{\{bb\}}_{\{\bar c\bar c\}}\to   \overline \Xi_{c}^-  \Xi_{bb}^{-} $ & $ -2 f_4$&
$T^{\{bb\}}_{\{\bar c\bar c\}}\to   \overline \Xi_{c}^{\prime-}  \Xi_{bb}^{-} $ & $ \sqrt{2} f_5$\\\hline
\hline
$T^{\{bb\}}_{\{\bar c\bar c\}}\to   \overline \Lambda_{c}^-  \Xi_{bb}^{-} $ & $ 2 f_4 \text{sC}$&
$T^{\{bb\}}_{\{\bar c\bar c\}}\to   \overline \Xi_{c}^-  \Omega_{bb}^{-} $ & $ -2 f_4 \text{sC}$\\\hline
\hline
$T^{\{bb\}}_{\{\bar c\bar c\}}\to   \overline \Sigma_{c}^{-}  \Xi_{bb}^{-} $ & $ -\sqrt{2} f_5 \text{sC}$&
$T^{\{bb\}}_{\{\bar c\bar c\}}\to   \overline \Xi_{c}^{\prime-}  \Omega_{bb}^{-} $ & $ \sqrt{2} f_5 \text{sC}$\\\hline
\hline
$T^{\{bb\}}_{\{\bar c\bar c\}}\to   \overline \Lambda_{c}^-  \Omega_{bb}^{-} $ & $ -2 f_4 \text{sC}^2$&
$T^{\{bb\}}_{\{\bar c\bar c\}}\to   \overline \Sigma_{c}^{-}  \Omega_{bb}^{-} $ & $ \sqrt{2} f_5 \text{sC}^2$\\\hline
\hline
\end{tabular}
\end{table}
\section{Golden Decay Channels}
\label{sec:golden_channels}
In this section, we will discuss the golden channels to reconstruct the $T^{\{bb\}}_{\{\bar c\bar c\}}$. Our previous  classifications   are mainly  based on the CKM elements. In principle, the amplitudes of b-quark decay transitions such as $b\to c\ell^-   \bar \nu_{\ell}$, $b\to c\bar c s$ and $b\to c\bar u d$ will receive the largest contribution as $V_{cb}\sim 10^{-2}$. For the $\bar c$-quark decay, the  $\bar c\to \bar sd \bar u$ and $\bar c\to \bar s\ell^-   \bar \nu_{\ell}$ transition  has  the largest decay widths as $V_{cs}^*\sim 1$. In our analysis, the final meson can  be replaced by its corresponding counterpart with the same quark constituent but with the different $J^{PC}$ quantum numbers. For instance, one can replace a $\overline K^0$ by $\overline K^{*0}$.

Following the criteria~\cite{Li:2018bkh}, we can obtain the golden decay channels in Table~\ref{tab:Xb6_golden_meson_c}.
\begin{itemize}
\item Branching fractions:  For $\bar c$-quark decays, one should choose the corresponding channels with the transition of $\bar c\to \bar sd \bar u$ or $\bar c\to \bar s\ell^-   \bar \nu_{\ell}$, while for $b$-quark decays, the process with the quark level transition  $b\to c\ell^-   \bar \nu_{\ell}$ or $b\to c\bar c s$ or $b\to c\bar u d$ should be chosen.

\item Detection efficiency: At hadron colliders like LHC,  charged particles have higher rates to be detected than neutral states. So we will remove the channels with the final states $\pi^0$, $\eta$, $\phi$, $n$, $\rho^{\pm}(\to \pi^{\pm}\pi^0$), $K^{*\pm}(\to K^{\pm}\pi^0$) and $\omega$, but keep the modes with $\pi^\pm, K^0(\to \pi^+\pi^-), \rho^0(\to \pi^+\pi^-)$.

\end{itemize}
\begin{table}
 \caption{Cabibbo allowed $T^{\{bb\}}_{\{\bar c\bar c\}}$ $\bar c$-quark and b-quark decays respectively.  }\label{tab:Xb6_golden_meson_c}\begin{tabular}{|c  c   c   c c|}\hline\hline
$T^{\{bb\}}_{\{\bar c\bar c\}}\to \overline B^0_s  B_c^- l^-\bar\nu$&
$T^{\{bb\}}_{\{\bar c\bar c\}}\to   B^-  \overline B^0_s  D^- $&
$T^{\{bb\}}_{\{\bar c\bar c\}}\to   B^-  K^0   B_c^- $&
$T^{\{bb\}}_{\{\bar c\bar c\}}\to   \overline B^0_s  \pi^-   B_c^- $ &
\\
$T^{\{bb\}}_{\{\bar c\bar c\}}\to   \overline \Xi_{c}^-  \Xi_{bb}^{-} $&
$T^{\{bb\}}_{\{\bar c\bar c\}}\to   \overline \Xi_{c}^{\prime-}  \Xi_{bb}^{-} $& & &\\
\hline\hline
$T^{\{bb\}}_{\{\bar c\bar c\}}\to B_c l^-\bar\nu $ & 
$T^{\{bb\}}_{\{\bar c\bar c\}}\to   \pi^-   B_c^- $ &
$T^{\{bb\}}_{\{\bar c\bar c\}}\to   B^-  D^- $&
$T^{\{bb\}}_{\{\bar c\bar c\}}\to    D^-_s  B_c^- $&\\
$T^{\{bb\}}_{\{\bar c\bar c\}}\to B_c J/\psi l^-\bar\nu $ &
$T^{\{bb\}}_{\{\bar c\bar c\}}\to \overline D^0  B^- l^-\bar\nu $ &
$T^{\{bb\}}_{\{\bar c\bar c\}}\to D^-  \overline B^0 l^-\bar\nu $ &
$T^{\{bb\}}_{\{\bar c\bar c\}}\to  D^-_s  \overline B^0_s l^-\bar\nu $ & \\

$T^{\{bb\}}_{\{\bar c\bar c\}}\to    D^-_s  B_c^-  J/\psi $ &
$T^{\{bb\}}_{\{\bar c\bar c\}}\to   \overline D^0   D^-_s  B^- $ &
$T^{\{bb\}}_{\{\bar c\bar c\}}\to   D^-   D^-_s  \overline B^0 $ &
$T^{\{bb\}}_{\{\bar c\bar c\}}\to    D^-_s   D^-_s  \overline B^0_s $ &\\
$T^{\{bb\}}_{\{\bar c\bar c\}}\to   \overline D^0  B_c^-  K^-  $ &
$T^{\{bb\}}_{\{\bar c\bar c\}}\to   D^-  B_c^-  \overline K^0  $ &
$T^{\{bb\}}_{\{\bar c\bar c\}}\to   \pi^-   J/\psi  B_c^- $ &
$T^{\{bb\}}_{\{\bar c\bar c\}}\to    D^0  D^-  B_c^- $ &\\
$T^{\{bb\}}_{\{\bar c\bar c\}}\to   B^-  D^-  J/\psi $ &
$T^{\{bb\}}_{\{\bar c\bar c\}}\to   B^-  \overline D^0  \pi^-  $ &
$T^{\{bb\}}_{\{\bar c\bar c\}}\to   B^-   D^-_s  K^0  $ &
$T^{\{bb\}}_{\{\bar c\bar c\}}\to   \overline B^0  D^-  \pi^-  $ &\\
$T^{\{bb\}}_{\{\bar c\bar c\}}\to   \overline B^0_s  D^-  K^-  $ &
$T^{\{bb\}}_{\{\bar c\bar c\}}\to   \overline B^0_s   D^-_s  \pi^-  $ &
$T^{\{bb\}}_{\{\bar c\bar c\}}\to   K^0   K^-   B_c^- $ &
$T^{\{bb\}}_{\{\bar c\bar c\}}\to   \overline D^0   D^-_s  B_c^- $ &\\

$T^{\{bb\}}_{\{\bar c\bar c\}}\to   \overline \Xi_{cc}^{--}  \Xi_b^0 $&
$T^{\{bb\}}_{\{\bar c\bar c\}}\to   \overline \Xi_{cc}^{-}  \Xi_b^- $&
$T^{\{bb\}}_{\{\bar c\bar c\}}\to   \overline \Xi_{cc}^{--}  \Xi_{b}^{\prime0} $&
$T^{\{bb\}}_{\{\bar c\bar c\}}\to   \overline \Xi_{cc}^{-}  \Xi_{b}^{\prime-} $&\\
$T^{\{bb\}}_{\{\bar c\bar c\}}\to   \overline \Omega_{cc}^{-}  \Omega_{b}^{-} $&
$T^{\{bb\}}_{\{\bar c\bar c\}}\to   \Omega_{bc}^{0}  \overline \Omega_{ccc}^{--} $&
$T^{\{bb\}}_{\{\bar c\bar c\}}\to   \overline \Xi_{c}^-  \Xi_b^- $&
$T^{\{bb\}}_{\{\bar c\bar c\}}\to   \overline \Lambda_{c}^-  \Sigma_{b}^{-} $&\\
$T^{\{bb\}}_{\{\bar c\bar c\}}\to   \overline \Sigma_{c}^{--}  \Lambda_b^0 $&
$T^{\{bb\}}_{\{\bar c\bar c\}}\to   \overline \Xi_{c}^-  \Xi_{b}^{\prime-} $&
$T^{\{bb\}}_{\{\bar c\bar c\}}\to   \overline \Xi_{c}^{\prime-}  \Xi_b^- $&
$T^{\{bb\}}_{\{\bar c\bar c\}}\to   \overline \Sigma_{c}^{--}  \Sigma_{b}^{0} $&\\
$T^{\{bb\}}_{\{\bar c\bar c\}}\to   \overline \Sigma_{c}^{-}  \Sigma_{b}^{-} $&
$T^{\{bb\}}_{\{\bar c\bar c\}}\to   \overline \Xi_{c}^{\prime-}  \Xi_{b}^{\prime-} $&
$T^{\{bb\}}_{\{\bar c\bar c\}}\to   \overline \Xi_{cc}^{--}  \Xi_{bc}^{0} $&
$T^{\{bb\}}_{\{\bar c\bar c\}}\to   \Xi_b^0  \overline \Omega_{ccc}^{--}$&\\
$T^{\{bb\}}_{\{\bar c\bar c\}}\to   \Xi_{b}^{\prime0}  \overline \Omega_{ccc}^{--}$& & & &\\
\hline\hline
\end{tabular}
\end{table}

\section{Conclusions}
\label{sec:conclusions}
Although many charmonium-like and bottomonium-like states have been found on experimental side, our current knowledge on  hadron exotics is still far from mature.  The understanding on the hadron spectroscopy can be deepen by the study of exotic states of new categories. In this direction,
the fully-heavy tetraquark $T^{\{bb\}}_{\{\bar c\bar c\}}$ are of great interest. In this paper, we have discussed the lifetime and the weak decays. From our calculation, the lifetime of $T^{\{bb\}}_{\{\bar c\bar c\}}$ is found about $0.1-0.3$ ps. 
We have systematically discussed the possible weak decay modes, such as two- or three-body mesonic decays and two-body baryonic decays.  Finally, we  have collected  the golden channels  of $T^{\{bb\}}_{\{\bar c\bar c\}}$ with the largest branching fraction and experimental detector efficiency.
Our results for the lifetime and golden channels are helpful to search for the fully-heavy tetraquark in future experiments.

\section*{Acknowledgments}

 This work is
supported in part by   National Natural Science Foundation of
China under Grant No.~11575110, 11655002, 11675091, 11735010, and 11835015, by Natural Science Foundation of Shanghai under Grant  No.~15DZ2272100,  by Key
Laboratory for Particle Physics, Astrophysics and Cosmology,
Ministry of Education.

\end{document}